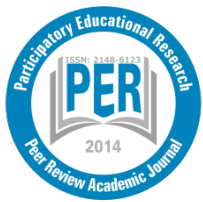

# Building Journal Impact Factor Quartile into the Assessment of Academic Performance: A Case Study

## Keziban Orbay

*Department of Mathematics and Science Education, Faculty of Education, University of Amasya, Turkey. ORCID: 0000-0002-7642-4139*

## Ruben Miranda

*Department of Chemical and Materials Engineering, Faculty of Chemistry, Complutense University of Madrid, Madrid, Spain. ORCID: 0000-0002-3675-247X*

## Metin Orbay[*]

*Department of Mathematics and Science Education, Faculty of Education, University of Amasya, Turkey. ORCID: 0000-0002-2609-1874*

This study aims to provide information about the Q Concept defined as the division of journal impact factors into quartiles based on given field categories so that the disadvantages resulting from the direct use of journal impact factors can be eliminated. While the number of "Original articles published in the Web of Science (WoS) database-indexed journals like SCI, SSCI and A&HCI" is an important indicator for research assessment in Turkey, neither the journal impact factors nor the Q Concept of these papers have been taken into account. Present study analyzes the scientific production of the Amasya University researchers in journals indexed in WoS database in the period 2014-2018 using the Q concept. The share of publications by Q category journals as well as the average citations received by the works from Amasya University were compared to the average situation in Turkey and other different countries in the world. Results indicate that the articles published by Amasya University researchers were mostly published in low impact factor journals (Q4 journals) (36.49%), in fact, only a small share of papers were published in high impact journals (14.32% in Q1 journals). The share of papers published in low impact journals by researchers from Amasya University is higher than the Turkish average and much higher than the scientific leading countries. The average citations received by papers published in Q1 journals was around six times higher than papers published in Q4 journals (8.92 vs. 1.56), thus papers published in Q1 journals received 30.02% citations despite only 14.32% of the papers was published in these journals. The share of papers published which were never cited in WoS was 27.48%, increasing from 9.68% in Q1 to almost half (48.10%) in Q4. The study concludes with some suggestions on how and where the Q Concept can be used.

[*] Corresponding author: metin.orbay@amasya.edu.tr

**Introduction**

Original articles published in the international scholarly journals and the citations that they receive from other researchers are among the fundamental criteria for the assessment of academic performance. Rather than the quantity, the quality of the articles (criteria such as the journal impact factor, whether articles are turned into patents, the citations they receive) and the interdisciplinary differences must be taken into account.

Nowadays, articles published in the journals indexed on the *Science Citation Index (SCI)*, *Social Sciences Citation Index (SSCI)* and *Arts & Humanities Citation Index (A&HCI)* in the Web of Science (WoS) database are predominantly used in the assessment of academic performance. The WoS database abstracts and indexes many journals in the fields of Science, Social Sciences, Arts, and Humanities, while calculates some citation indicators such as the journal impact factor (JIF). As is well-known, the journal impact factor is defined as the number of citations in the current year to items published in the previous two years, divided by the total number of scholarly citable items published in those same two years (Garfield, 1994). The fact that an article is published in a journal indexed in the WoS database and with a high impact factor is an indicator of a high quality, but cannot be guaranteed (Garfield, 1972). It doesn't necessarily mean that all the articles in the journals indexed in these databases are of higher quality than those published in journals which are not indexed. However, it is also true that researchers are actually urged to publish in journals indexed at well recognized scientific databases such as WoS or Scopus and especially in those journals with the highest impact factors in their fields. This and other similar cases have already been discussed in the San Francisco Declaration on Research Assessment (DORA, 2018).

The WoS database consists of over 9200 journals in 178 categories in SCI, over 3400 journals in 58 categories in SSCI, and over 1800 journals in 28 categories in A&HCI. It should be remembered that some journals can be indexed in more than one subject category: two, three and even four categories. Most of the cases these categories are of the same database, however, there are a few cases of journals indexed at the same time in two different databases, i.e. SCI and SSCI (Clarivate Analytics, 2019a).

***On JIF Quartile (Q)***

It might be misleading to look at the journal impact factors only in assessing the articles in the journals indexed in the WoS database even though they are in the same index. Variables like the interdisciplinary differences of the studies, the national and international dimensions studied, the popularity of the research problem, the number of researchers involved, and the number of journals in the field mean very different citation patterns (Bornmann & Marx, 2014; Waltman, 2016; Waltman & van Eck, 2019). For example, the journals "*Mediterranean Journal of Mathematics*" (JIF-2018: 1.181) and "*Future Oncology*" (JIF-2018: 2.279) are both indexed in SCI and the latter is twice as much as the former in terms of the impact factor. At first sight, the journal with a higher impact factor may seem to be of higher quality. However, comparing the JIF of journals from different fields is very questionable. The impact factors for the journals are calculated by finding an average number of citations in all the articles. Even the journal impact factor is the most used bibliometric method, it is largely depended on the skewness in the distribution of citations, the interdisciplinary differences, the fact that it takes into account the non-research articles, and editorial policies (for a detailed discussion see also: Archambault & Larivière, 2009; Bornmann & Williams, 2017a, 2017b; Callaway, 2016; Hammarfelt & Rushforth, 2017; Peters, 2017; Seglen, 1997; Waltman & Traag, 2017). Indeed, it could be a "mortal sin" to assess articles by looking at the impact factors of the journals (as quoted by Van

Noorden from Van Rann, 2010). While all this is under debate, there is one thing we should remember: In countries where nepotism prevails, bibliometric tools can provide objective and consistent assessment for researchers; thus, it would be possible to reach fast, fair and transparent decisions (Tang & Hu, 2018). The anxiety that the peer reviews in Turkey can be subjective and the idea that the journals in the WoS database accept articles after a review by experts in their fields are what directs us towards this type of assessment. In fact, such a tendency can be depicted as trapped into a "fatal attraction" of bibliometric methods (Van Raan, 2005). It must be remembered that despite the fact that the journal impact factor has some disadvantages, bibliometric methods produce correct outcomes in many respects, and are easy to calculate, and most importantly, it hasn't been replaced by another concrete method yet (Tregoning, 2018).

Taking into account these limitations, the WoS database has started to use a different way of assessment resulting from the interdisciplinary differences, to partially eliminate the "mortal sin", instead of assessing inter-journal impact factors in a wholesale fashion. Recently, this new assessment has got into the related literature as the Q Concept (Clarivate Analytics, 2019b). It may remind us of the concepts like *quantity* and *quality* that we often hear. It is, in fact, short of Quartile and what is intended is the analytical classification of the quality of journals (hence, the quality of articles). Now let's sum up how the quartile is calculated.

The journals indexed in the WoS database are classified under 254 key categories. Some journals fall under more than one category depending on their scopes and each category falls into subcategories. The journals under the same category are ranked from the highest impact factor to the lowest. Take the journal "*Computers & Mathematics with Applications*" indexed in SCI as an example. This journal falls under the "*Mathematics-Applied*" category. There is an aggregate of 254 journals under this category in the WoS database. In the first place, these journals are ranked from the highest impact factor to the lowest and divided into its quartiles. Those in the first quartile are classified as Q1 (within the top 25% of JIF among a certain category), the ones in the second quartile as Q2, those in the third quartile as Q3, and finally those in the fourth quartile as Q4-within the lowest 25% of JIF among a certain category (Clarivate Analytics, 2019b). Consequently, the journal "*Computers & Mathematics with Applications*" is ranked as 18[th] of all the 254 journals and is classified as Q1 as it is in the first quartile. Therefore, the two journals under the same category as the journal "*Mathematics-Applied*" can be compared more consistently in terms of the quality of articles in accordance with the Q category. We can unpack this with a concrete example. Imagine that there were two applications for an academic position. Let's look at the files of two candidates and consider the two different studies under the category "*Plant Sciences*". One of them got his research published in the journal "*Trends in Plant Science*" (JIF-2018: 14.006-Q1) and the other candidate had his work published in the journal "*Acta Botanica Mexicana*" (JIF-2018: 0.661-Q4). How possible is it to assess the two articles as equal in a wholesale fashion just because these two journals are both indexed in SCI. Firstly, it will be helpful to remember the disadvantages of the journal impact factors mentioned in the previous section before we attempt to answer this. Whatever the case, journals must be ensure the articles are reviewed by field experts irrespective of the journal impact factors. However, as it has been pointed out earlier, the anxiety that the peer reviews can be subjective and the idea that the journals in the WoS database accept articles after a review by experts in their fields are what directs us towards this type of assessment.

The journal impact factors may change for a variety of reasons ranging from the increasing number of journals to the selectiveness resulting from journal policies, the fact that they attract good studies, the possibility of reaching a wide audience and the international interest in the studies published (Bornmann & Marx, 2014; Waltman, 2016; Waltman & van Eck, 2019). This may change their Q categories, as well (Clarivate Analytics, 2019b).

The Q Concept, which is the division of journal impact factors into quartiles in a field, was embraced in the literature in a short time and used in many studies (e.g., Alverez et al. 2014; Bornmann & Marx, 2014; Chinchilla-Rodriguez et al. 2015; Miranda & Garcia-Carpintero, 2019; Liu, Hu & Gu, 2016; Tang, Shapira & Youtie, 2015; Zhaou & Lv, 2015).

Similar case about using Q category is seen in ranking of the universities. The university ranking systems consist of a variety of indicators in addition to the number of articles published in journals indexed internationally and the number of citations to these articles. The data from the university ranking systems (URAP-Turkey, THE-United Kingdom, Leiden-Netharlands, ARWU-China, US News-USA, etc.) by different organizations is closely watched by the internal as well as the external stakeholders of the universities even though they are criticized for their suitability of criteria that they use. The studies in the journals under the Q4 journals are often excluded in the assessments (URAP, 2019a).

It is one of the fundamental responsibilities of the scholarly people to conduct scholarly research, share the findings with the scholarly community and contribute to the field. As a natural result of the studies, they are appointed or promoted to higher academic positions. The bibliometric methods are used in the assessment of academic performance in Turkey such as the appointment and promotion to the position of faculty member, academic incentive programs, and the eligibility for the position of associate professor (Miranda & Garcia-Carpintero, 2018). Original articles published in the WoS database-indexed journals like SCI, SSCI and A&HCI play an important part in the criteria for the appointment and promotion to the position of faculty member, and the eligibility for the position of associate professor. (HEC, 2019; HIUC, 2019). However, the Q concept developed to deal with the disadvantages of using the journal impact factor(s) directly is not often taken into account. The JIF quartile is worth studying in the bibliometric assessment of academic performance.

**Research Aims**

The objective of this study is to investigate the scientific production from Amasya University (a Turkish state university-established in 2006) published between 2014 and 2018 in journals indexed in the WoS database taking into account the JIF quartiles. The research questions are as below:
   a) What is the share of the journals in which the articles were published by the Q category like?
   b) What is the share like in comparison to Turkey and the world?
   c) What kind of relationships are there between the Q categories of the journals and the citations made to the Amasya University-based articles?
   d) What are the journals that the researchers prefer most? What kind of relationships are there between the Q categories of the journals and the average citations received?

**Methodology**

The data set of the study consists of the Amasya University-based papers published between the years 2014-2018 in the journals indexed in WoS database. They were obtained using the "Organizations-Enhanced" tool from WoS. With the help of the bibliometric methods, first defined by Pritchard (1969) as *"the application of mathematical and statistical methods to* *books and other media of communication"*, the data from the *WoS* database was interpretered after that had been turned into tables and figures using the values of frequency and the distribution of percentages in accordance with the basic objectives set. The findings were summarized in general and presented by interpreting them in such a way that the audience could understand easily. As we have pointed out in the introduction, journals may go into more than one JIF quartile. There are two options that we can turn to so that we can avoid the double-counting problem: allocating the journal to the higher quartile, which is called "the optimistic mode" and if the journal is allocated to the lower quartile, then it is called "the pessimistic mode" (Liu, Hu, & Gu, 2016). We have chosen to use the optimistic mode in our analysis of the data from this point onwards.

**Results and Discussion**

There is an aggregate of 498 publications from Amasya University researchers in the period 2014-2018 in journals indexed in WoS, which means 0.27% of total papers published by Turkish research organizations. The 498 papers published were distributed by databases as follows: 467 (93.77%) in SCI, 27 (4.02%) in SSCI and only 3 (0.60%) in A&HCI databases. Furthermore, 10 additional publications indexed simultaneous in two databases: 8 were indexed both in SCI and SSCI, one in SCI and A&HCI and other in SSCI and A&HCI.

The main document types are: articles (433, 86.95%), reviews (3, 0.60%), proceeding papers (39, 7.83%), meeting abstracts (13, 2.61%), and others (10, 2.01%). These values were obtained considering as "proceeding papers" all the papers doubled classified as "article" and "proceedings papers" (39 items) and as "book chapter" to the single paper classified simultaneously as "review" and "book chapter".

The three articles indexed in A&HCI were all published in Turkish journals in the years 2014, 2016, and 2017 and have not received any citations yet. Due to the journals indexed in A&HCI category has not JIF, they were excluded from the present study, focused on applying Q concept. Subsequent studies have focused on two document types (articles and reviews) either published in SCI or SSCI database, which amounts 433 papers. Table 1 shows the share of papers and the citations they received by database and JIF quartiles.

**Table 1**. The share of articles by the Q category and the citations they received.

| Index | Q1 | Q2 | Q3 | Q4 | Paper Count | Total Citations | Average Citations Per Paper |
|-------|----|----|----|----|-------------|-----------------|-----------------------------|
| SCI | 61 | 87 | 119 | 143 | 410 | 1725 | 4.20 |
| SSCI | 1 | 3 | 4 | 15 | 23 | 50 | 2.17 |
| **Total** | 62 | 90 | 123 | 158 | 433 | 1775 | 4.10 |

As can be seen in Table 1, the number of Amasya University-based articles indexed in the SSCI is much lower than those published in SCI (410 vs. 23) as well as their average citations (it is well known that papers published in Social Sciences are much lower cited than papers in Sciences). A 64.90% of total papers were published in Q3 and Q4 journals, however, there is a significant greater share of papers published in these journals in Social Sciences (82.61%) than in Sciences (63.90%).

Table 2 shows the share of the journals by the Q category in which an aggregate of 433 articles that come under articles and reviews were published.

**Table 2.** The classification and share of the articles by the Q category and year.

| Year | Q1 | Q2 | Q3 | Q4 | Q1 (%) | Q2 (%) | Q3 (%) | Q4 (%) |
|---|---|---|---|---|---|---|---|---|
| 2014 | 8 | 22 | 20 | 30 | 10.00 | 27.50 | 25.00 | 37.50 |
| 2015 | 9 | 10 | 14 | 22 | 16.36 | 18.18 | 25.46 | 40.00 |
| 2016 | 12 | 19 | 28 | 30 | 13.48 | 21.35 | 31.46 | 33.71 |
| 2017 | 13 | 23 | 36 | 40 | 11.61 | 20.53 | 32.15 | 35.71 |
| 2018 | 20 | 16 | 25 | 36 | 20.62 | 16.49 | 25.78 | 37.11 |
| Paper (% Paper) | 62 (*14.32*) | 90 (*20.78*) | 123 (*28.41*) | 158 (*36.49*) | | | | |
| Citations (% Citations) | 533 (*30.02*) | 503 (*28.34*) | 492 (*27.72*) | 247 (*13.92*) | | | | |

As shown in Table 2, the percentage of papers published in Q1 journals was only 14.32% while most papers were published in the Q4 journals (36.49%). Although the share of papers published in Q1 was 14.32%, they received 30.02% total citations from Amasya University-based articles received, which was an expected outcome because the impact factors in the Q1 journals are high, they are read by more researchers and receive more citations (Garfield, 2006; Miranda & Garcia-Carpintero, 2019). On the other hand, the papers published in Q4 received only 13.92% total citations despite a 36.49% of total papers were published in these journals. Such a case pointed out that Q4 journals were less cited and potentially less read. Meanwhile, the Amasya University-based articles have had 1775 citations so far. 128 of them are self-citations, which represents 7.21%. This can be regarded as positive.

Depending on the impact factors of the journals and the Q categories, the quantity and quality debate is pointed out in the URAP reports, the first national ranking system for Turkish universities (URAP, 2019b, 2019c). The reports indicate that despite the fact that the number of the Turkey-based articles increased every year, most Turkish universities relegated when compared to other countries both in the URAP rankings as well as the international rankings (URAP, 2019b, 2019c). The underlying reason is that the number of articles published in the journals with a very low or next to zero impact factor has increased recently. When the data in URAP reports and Table 2 are compared, 41.9% of the articles published world-wide in 2018 were in the Q1 journal whereas only 14.32% of the Amasya University-based articles appeared in the Q1 journals, which was half of the world average. On the other hand, only 13.15% of the world publications were published in the Q4 journals while the percentage of the Amasya University-based articles published in the Q4 journals was at about 37.11%. The main differences between Amasya University and Turkey average is the share of papers published in Q1, which was around double in some years, however, the share of papers in Q4 journals is very similar in the period considered. When compared to the world average, the share of papers published in Q1 journals by Amasya University is from 2 to 4 times lower, while the share of papers published in Q4 journals was between 2 and 3 times higher (Figure 1).

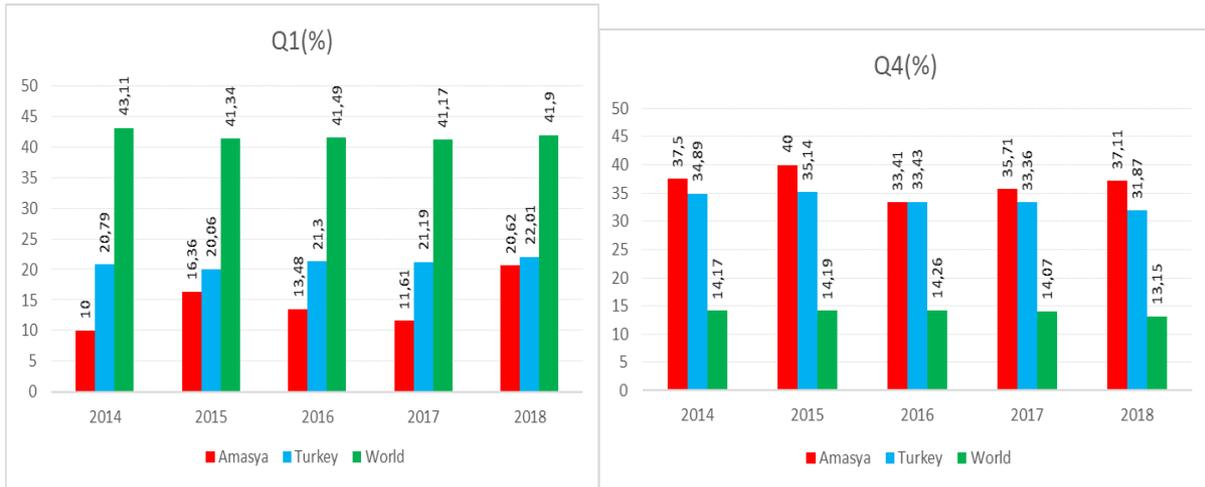

**Figure 1.a.** Comparison of the publication percentages of articles by the Q1 and Q4 journals related to years.

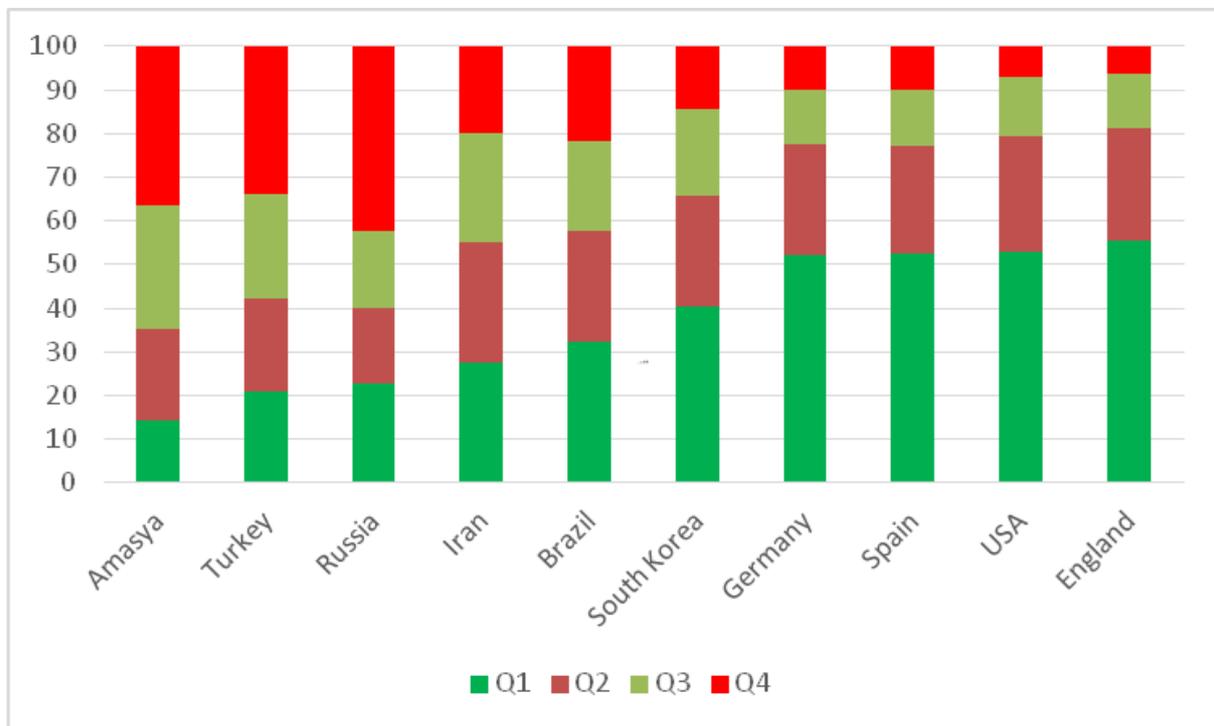

**Figure 1.b.** Comparison of the publication percentages of articles by the journal quartiles between 2014 and 2018.

Table 3 indicates the share of articles with no citations by the Q category and the year. As can be seen in Table 3, 119 out of the 433 papers published (27.48%) were not cited. The uncitedness increased largely by quartiles, from 9.68% in Q1, to 14.44% in Q2, 19.51% in Q3 and especially to 48.10% in Q4. As expected, the share of articles with no citations is greater at the recent years as these papers had less time to accumulate citations than an older ones. However, these articles still have the potential to receive citations in the future. Similarly, the journals in which the 10 most cited papers from Amasya University researchers, the Q categories and the number of citations each article has received. Seven of the 10 most cited articles were published in the Q1 journal, two articles were in the Q2 journal, and one article was in the Q3 journal. The data is supportive of the fact that the Q1 journals with a high impact factor are read more (Miranda & Garcia-Carpintero, 2019).

**Table 3.** The share of articles with no citations by the Q category and the year.

| Year | Q1 | Q2 | Q3 | Q4 | No Cited Paper (% papers) |
|---|---|---|---|---|---|
| 2014 | - | - | 1 | 8 | 9 (11.25%) |
| 2015 | - | - | 1 | 3 | 4 (7.27%) |
| 2016 | 1 | 1 | 2 | 14 | 18 (20.22%) |
| 2017 | 2 | 3 | 10 | 23 | 38 (33.93%) |
| 2018 | 3 | 9 | 10 | 28 | 50 (51.55%) |
| **2014-2018 (% papers)** | **6 (9.68%)** | **13 (14.44%)** | **24 (19.51%)** | **76 (48.10%)** | **119 (27.50%)** |

Table 4 shows the journals in which the researchers at Amasya University prefer the most, the field categories of these journals, and the average number of citations per article. All the data in the WoS database and in Table 4 indicate that the field categories focus rather on the fields of Physics, Chemistry, Biology, and Mathematics, which is directly related to the number of researchers employed by Amasya University.

**Table 4.** Some bibliometric values for the journals in which the researchers prefer most.

| Journal | Q | Category(s) | Paper Count | Average Citations Per Paper |
|---|---|---|---|---|
| Filomat (SCI) | 2 4 | Mathematics Mathematics-Applied | 15 | 1.60 |
| Journal of Molecular Structure (SCI) | 3 | Chemistry-Physical | 14 | 6.50 |
| Fresenius Environmental Bulletin (SCI) | 4 | Environmental Sciences | 11 | 1.00 |
| Crystallography Reports (SCI) | 4 | Crystallagraphy | 10 | 0.40 |
| Spectrochimica Acta Part A: Molecular and Biomolecular Spectroscopy (SCI) | 1 | Spectroscopy | 8 | 13.75 |
| Bangladesh Journal of Botany (SCI) | 4 | Plant Sciences | 6 | 0.83 |
| Journal of Intelligent Fuzy Systems (SCI) | 3 | Computer Science-Artifical Intelligence | 5 | 11.20 |
| Miskolc Mathematical Notes (SCI) | 4 | Mathematics | 5 | 1.80 |
| Biomedical Research India (SCI) | 4 | Medicine Research-Experimental | 5 | 1.20 |
| Materiali in Tehnologije (SCI) | 4 | Materials Science-Multidisciplinary | 5 | 1.00 |

The WoS database was the sole provider in the indexing of citations up until the 2000s. Recently, it has expanded its operations to include locally recognized new journals in its indexing after the arrival of some other platforms like Scopus and Google Scholar. Since the year 2015, the Emerging Sources Citation Index (ESCI) has been active in the WoS database. There are over 7800 journals (254 categories) indexed in the ESCI whose criteria of eligibility seem to be less strict. The underlying reason for this is to make known the journals that are scientifically significant but are unknown to the scholarly community in the world yet (Testa, 2016; Clarivate Analytics, 2019c).

There are Amasya University-based 126 articles and three reviews published in the journals indexed in the *ESCI* between the years 2015-2018 (Figure 2).

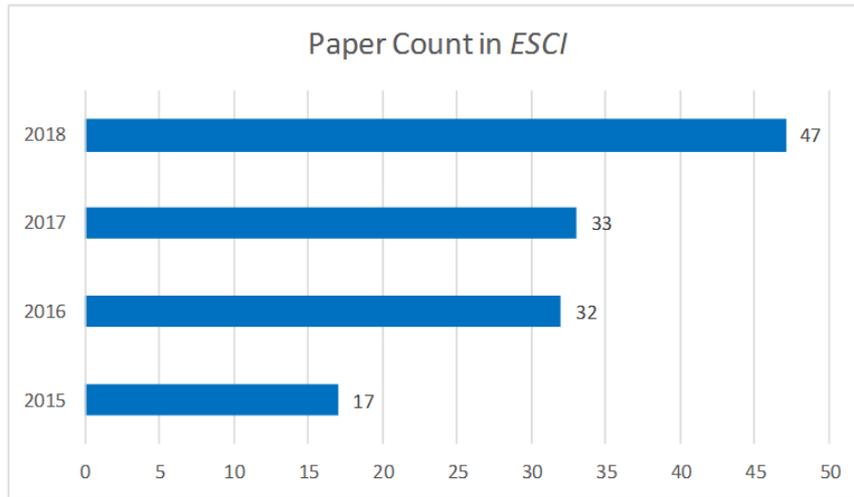

**Figure 2.** Articles published in the journals indexed in the *ESCI* between the years 2015-2018.

A 60.3% of the articles indexed in the ESCI was published in Turkey-based journals. This is consistent with the ESCI principle of giving priority to locally significant journals. On the other hand, it is advisable to remember the phrase labelling the Turkey-based articles in the journals indexed in the WoS database as the *locally international, but internationally local journals* (Tonta, 2017a). This kind of labelling is firstly used for the international journals in which the number of international researchers and audience is thin, and which can receive no citations from the articles in the other international journals (Pajić & Jevremov, 2014). It must be borne in mind that some of the Turkey-based articles in the journals indexed in the WoS database exhibit local characteristics in terms of researchers and have a lot of self-citations result in dropped from Web of Science (Doğan, Dhyi, & Al, 2018).

**Conclusions**

Academia means people open to new knowledge and change and who keep improving themselves. Therefore, universities have always been organizations that open up new doors in society and push the envelope during the course of history. They are dynamic in that they are in search of *quality* in their academic activities they carry out, *innovation* in the research and development they do, and *stakeholder satisfaction* in the services they deliver. That is why they always compete against each other. The social dynamics of change like globalization, internationalization, and economies of knowledge that have an in/direct impact on universities constantly shape the areas of activities for and the moulds of expectations towards universities. It seems that the JIF quartile will be under debate and be used by bearing in mind the fact that each assessment method has its own advantages and disadvantages in itself among the other methods of evaluation based on quality and performance in the community of the higher education.

The quantitative values like the number of articles published must be supported together with the qualitative elements in the assessment of academic performance. The bibliometric tools like JIF and JIF quartile and the journals are used as qualitative elements for publications. While the entry "*Original articles published in the WoS database-indexed journals like SCI, SSCI and A&HCI*" is considered important in the eligibly criteria for the position of associate professor and in the appointment and promotion to the position of faculty member in most of the universities in Turkey, neither the JIF nor the JIF quartile values are taken into account yet, as occurs in many European countries (Abambres, Ribeiro, Sousa, & Lantsoght, 2018) On the other hand, researchers are paid a sum of money for the articles published in the journals indexed in the WoS database under the TÜBİTAK Incentive Program for International Scientific Publications (TÜBİTAK, 2019). While a variety of practices such as JIF and JIF quartiles are used to encourage the quality of publications under this program, it is not coherent that they are not built into the appointment and promotion to the position of faculty member and the eligibility for the position of associate professor (HEC, 2019; HIUC, 2019). It can be said that the fact that articles are assessed in a wholesale fashion, often without taking JIF and JIF quartile into account are among the reasons why researchers often prefer getting their research published in the Q4 journals. Necessary arrangements must be made by avoiding assessment done in a wholesale fashion and by observing the interdisciplinary differences in the eligibility criteria for the position of associate professor and in the appointment and promotion to the position of faculty member; otherwise, it wouldn't be possible to stop articles from being published just to get an academic position even if they were indexed in the WoS.

Recently, the JIF quartile has begun to be taken into account by some of the universities in the appointment and promotion to the position of faculty member. For instance, Ege University (a Turkish state university-established in 1955) rates articles by the JIF quartile in the assessment of academic performance as Q1=50, Q2=40, Q3=30 and Q4=20, respectively (HEC, 2019). Sabancı University (a Turkish foundation university-established in 1994) has the eligibility criteria of having to have at least 3, 8, 16 articles in the Q1 or Q2 quartiles for the position of Assistant Professor, Associate Professor and Professor respectively at the Faculty of Engineering and Natural Sciences (HEC, 2019). This seems promising for the future, but these should be done by taking the opinions of the field experts. It would be beneficial to take into account the incentive programs and related studies about the behavior of Turkish academicians (Asan & Aslan, 2020; Demir, 2018a, 2018b; Tonta, 2017a, 2017b, 2017c; Tonta & Akbulut, 2019; Yuret, 2017). Each university is expected to put similar practices into effect by taking into account the academic staff, the physical infrastructure, and the mission and vision in particular on the basis of interdisciplinary differences. It will be one of the important steps to be taken to achieve the quality of assessment if certain criteria are sought out about the faculty members (the panel for the assessment of the position of Professor and Associate Professor, or the PhD Dissertation Defense) who will do the assessment of academic performance.

Consequently, it is positive that the assessment of academic performance has been under debate by the community of the higher education recently in Turkey. It is a step forward to seek out the eligibility criteria of getting your articles published in the journals indexed in the internationally recognized databases by taking into account the interdisciplinary differences. Bibliometric methods like impact factors, h-index values and article citations will be discussed in the future. We must remember the phrase by sociologist William Bruce Cameron while assessing the academic performance in terms of some figures and numbers. As he put it: "*Not everything that counts can be counted, and not everything that can be counted counts.*" (Cameron, 1963). In Turkey, we seem to be leading an academic life along with the "fatal attraction" of bibliometric methods-until we find a better way (Tregoning, 2018). Al and Soysal (2014) feels that *"The war of academia with citation indexes"* will go on for a long time in Turkey. With the DORA on our minds, bibliometric tools like the JIF quartile will offer benefits. In the meantime, it is absolutely necessary to take advantage of the accumulated knowledge by field experts on studies concerning the assessment of academic performance.

**Limitations and Future Study**

We acknowledge that this study has a few limitations. First, bibliometric indicators based on citation number are time-dependent indicators and can change over time. Second, the year selection and the small sample size limit the generalizability of our findings. Future studies could extend the sample size and it would be useful to use interviews and other primary data methods to further probe why the researchers primarily choose Q4 journals.